\documentclass[manuscript]{acmart}

\copyrightyear{2020}
\acmYear{2020}
\setcopyright{None}
\acmConference[CARS '20]{CARS 2.0: Workshop on Context-Aware Recommender Systems}{Sept 25, 2020}{}
\acmBooktitle{CARS '20: CARS 2.0: Workshop on Context-Aware Recommender Systems, Sept 25, 2020}
\acmPrice{}
\acmDOI{}
\acmISBN{}
\fancypagestyle{firstpagestyle}{%
\fancyhead[L]{}%
\fancyfoot[RO,LE]{\if@ACM@printfolios\small\thepage\fi}%
\fancyfoot[RE,LO]{\footnotesize Copyright held by the authors}%
}

\renewcommand\footnotetextcopyrightpermission[1]{} 

\makeatletter
\let\@authorsaddresses\@empty
\makeatother

\usepackage{url}
\usepackage{enumitem}
\usepackage{subfigure}
\usepackage{multirow}

\usepackage{color}
\newcommand{\myblue}[1]{\textcolor[rgb]{0.00,0.00,0.00}{{#1}}}
\newcommand{\bb}[1]{\textcolor[rgb]{0.00,0.00,0.00}{#1}}

\usepackage{soul}
\setstcolor{red}
\newcommand{\myitem}[1]{\vspace{0.25\baselineskip}\noindent\textbf{#1}}

\newcommand{\mycustomcolumnwidth}{0.8\columnwidth}
\newcommand{\mycustomcolumnwidthA}{0.7\columnwidth} 
\newcommand{\mycustomcolumnwidthB}{0.6\columnwidth} 
\newcommand{\mycustomwidth}{0.46\linewidth}

\definecolor{orange}{rgb}{1,0.5,0}

\newcommand{\removeTextForCARSsubmission}[1]{}

\begin{document}

\title{Towards QoS-Aware Recommendations}

        


\thanks{This research is co-financed by Greece and the European Union (European Social Fund- ESF) through the Operational Programme ``Human Resources Development, Education and Lifelong Learning'' in the context of the project ``Reinforcement of Postdoctoral Researchers - 2nd Cycle'' (MIS-5033021), implemented by the State Scholarships Foundation (IKY).   This project was also partially sponsored by CAPES, CNPq and FAPERJ, through grants  E-26/203.215/2017 and E-26/211.144/2019.
}

\author{Pavlos Sermpezis} \affiliation{Aristotle University of Thessaloniki, Greece}
\author{Savvas Kastanakis} \affiliation{FORTH-ICS, Heraklion, Greece}
\author{Jo\~ao Ismael Pinheiro} 
\author{Felipe Assis} 
\author{Mateus~Nogueira} 
\author{Daniel Menasch\'e} \affiliation{UFRJ, Rio de Janeiro, Brazil}
\author{Thrasyvoulos Spyropoulos} \affiliation{EURECOM, Sophia-Antipolis, France}

\renewcommand{\shortauthors}{P. Sermpezis \textit{et al.}}

%
\begin{abstract}
In this paper we propose that recommendation systems (RSs) for multimedia services should be ``QoS-aware'', i.e., take into account the expected QoS with which a content can be delivered, to increase the user satisfaction. Network-aware recommendations have been very recently proposed as a promising solution to improve network performance. However, the idea of QoS-aware RSs has been studied from the network perspective. Its feasibility and performance performance advantages for the content-provider or user perspective have only been speculated. Hence, in this paper we aim to provide initial answers for the feasibility of the concept of QoS-aware RS, by investigating its impact on real user experience. To this end, we conduct experiments with real users on a testbed, and present initial experimental results. Our analysis demonstrates the potential of the idea: QoS-aware RSs could be beneficial for both the users (better experience) \emph{and} content providers (higher user engagement). Moreover, based on the collected dataset, we build statistical models to (i) predict the user experience as a function of QoS, relevance of recommendations (QoR) and user interest, and (ii) provide useful insights for the design of QoS-aware RSs. We believe that our study is an important first step towards QoS-aware recommendations, by providing experimental evidence for their feasibility and benefits, and can help open a future research direction.


\end{abstract}

\maketitle

\section{Introduction}\label{sec:intro}
Multimedia delivery services, such as online video (e.g., YouTube, Netflix, Hulu), audio (e.g., Spotify, Deezer), live streaming for gaming (e.g., Twitch.tv), and content over social media (e.g., Facebook), use recommendation systems (RSs) to best satisfy users, and/or maximize their engagement. Recommendations are mainly based on user interests, history, content/user similarity scores, etc., and may also take into account the \textit{context} of a viewing session (e.g., location, time, type of device, user activity, environment)\myblue{~\cite{villegas2018characterizing,agagu2018context, adomavicius2011context, baltrunas2011matrix, nguyen2014gaussian, pagano2016contextual, almutairi2017context, wang2017context}}; for instance, a content might be relevant to a user at a given time of the day and/or a given location, but not at another occasion.


Similarly to other context parameters, the \textit{quality of service (QoS)} with which a content is delivered to the users may affect their experience (\textit{QoE}) or even make them abandon the service~\cite{conviva2015,ericsson2018,krishnan2013video,nam2016qoe,plakia2019should}. For example, a user may be interested in a video, but if there are a lot of re-bufferings (e.g., due to an overloaded video server), she may lose her interest in watching it. In another example, a user may regularly prefer to watch HD sports videos during her commute, but if she knew that at a part of the trip the only available video streaming quality is 144p (e.g., due to network congestion or poor signal), she may have preferred and enjoyed more a different type of video, e.g., a music video clip.

However, \textit{RSs for multimedia content services are currently agnostic to the \textit{QoS} (or the delivery network), and QoS has not been considered as a dimension of the ``context''}. This may result in poor user satisfaction or engagement. In fact, it is often observed that network conditions and topology affect the video streaming (bit-rate, latency, start-up delay, etc.) and may cause
annoying impairments such as re-buffering or changes in bit-rates~\cite{nam2016qoe,doan2019tracing,plakia2019should}. 
This concern is expected to be amplified in the future, as video delivery increasingly dominates network traffic, 
and larger portions of such content must be delivered over mobile networks, which will struggle to serve all content requests in high-QoS~\cite{ericsson2018,cisco2018}. 

\myitem{Towards a new paradigm: QoS-aware recommendations.} Motivated by the above arguments, \textit{we propose to explicitly take QoS into account in RSs for video and multimedia content delivery}. For example, if two videos have similar predicted interest to a user, but current network conditions suggest that one of them would likely suffer from poor quality streaming, the RS should recommend the high-QoS video. Our claim is that recommending content that can be delivered with high QoS (e.g., videos cached close to users) will lead to improved user experience and engagement.

The proposed paradigm finds also support in some very recent works in the area of communication networks, which have proposed to leverage recommendations for network optimization~\cite{sermpezis2018soft,chatzieleftheriou2019jointly,kastanakis2018cabaret,zhu2018coded,lin2018joint,giannakas2018show,song2018making,garetto2020similarity}. Initial results are very promising for the \textit{network performance}: nudging recommendations towards network-friendly content delivery can bring significant improvements to network capacity, congestion, or energy consumption. While these findings reveal a new potential, they have two shortcomings: (i) recommendation quality and user experience was a secondary goal, i.e., considered as ``something not to get too distorted'' while improving network performance; (ii) the impact of this recommendation nudging and potential (QoE) distortions is only speculated, but is not validated with real users.

\myitem{Goals and Contributions.} Our work aims to be the first step towards investigating the feasibility, trade-offs, and potential benefits of the QoS-aware recommendations paradigm. Our goal is to provide initial answers to some key questions that precede the adoption or the design of a QoS-aware RS: ``Should one design QoS-aware recommendation algorithms that explicitly target \emph{overall} user experience, rather than just looking for interesting content?''
. ``How would real users factor in QoS, content interest, and Quality of Recommendations (QoR) into their overall experience?''. While the earlier toy example, with two equally interesting contents, is straightforward, less obvious trade-offs arise in practice. For example, how should a recommender choose between a content with predicted QoS rating 3 (in a scale 1-5) and interest rating 4, and another with QoS 4 and interest 3? What about $\{4,3\}$ and $\{3,5\}$, respectively? Should an RS favor the content of higher interest to the user or of higher QoS? And does the answer change, depending on the scenario? 

In this paper, we provide \textit{initial experimental results and insights on the feasibility and design of QoS-aware RSs}. Specifically, we:
\begin{itemize}[nosep,leftmargin=*]
\item Build an experimental platform that emulates a video service, QoS impairments, and a simple QoS-aware RS (Section~\ref{sec:experiments}). 
\item Conduct experiments with users, collect their ratings, analyze results, and discuss interesting observations (Section~\ref{sec:results}).
\item Propose statistical models to quantify the user QoE as a function of the video QoS and the user interest in it (Section~\ref{sec:ml-model}).
\end{itemize}

\myitem{Key findings.} 
\bb{Our findings provide experimental evidence for the feasibility and benefits of QoS-aware recommendations; we pinpoint this as the main contribution of our study. In particular, they suggest that}:
\begin{itemize}[leftmargin=*,label={o},nosep]
\item Carefully nudging recommendations towards network friendly video delivery is not perceived as intrusive by users. This is a positive message for the feasibility of the recently proposed paradigm for joint network and RSs design
. 
\item QoS-aware recommendations can lead to a higher overall user satisfaction (on top of the expected network benefits).    
\item Recommending a high-QoS video is of equal (at least) importance to recommending an interesting 
video, for refraining users from abandoning a session. Hence, taking QoS into account in RSs can lead to higher revenues for the content provider as well.
\item \bb{To better satisfy a user, a QoS-aware RS should recommend contents that satisfy a minimum standard of \textit{both} QoS and user interest; then, among such contents, priority should be given to those of higher user interest.} 
\end{itemize}
In summary, we demonstrate 
that QoS-aware recommendations can benefit both the \textit{users} (better experience) and the \textit{RS} (higher user engagement), adding to the already shown gains for the \textit{network}. Moreover, we identify complex relationships between the quality of recommendations, the QoS, and the user satisfaction, which indicates that the design of QoS-aware RS is far from trivial and requires further investigation. To facilitate future research on this topic, we open-source the code of the experimental testbed and the collected dataset~\cite{cabaret-github}.

\section{Experimental Methodology}\label{sec:experiments}
We implemented an experimental platform to collect data from real users. The platform is built on top of YouTube: it streams videos through the YouTube service, and uses the YouTube API to retrieve recommended/related contents.



\subsection{Experiment Session}
We invited users to visit our platform and participate in our experiment. We first summarize here the steps of each experiment/session, and elaborate on some key steps subsequently. 

\myitem{Step 1:} The user enters the platform and is requested to select from a list her preferred \textit{region} (e.g., U.S., India, Brazil;~\cite{youtube-api-regions}). \\
\textbf{Step 2:} Then, she is redirected to a page with instructions about the experiment, and is asked to start the viewing session by selecting a video from a list of $20$ \myblue{trending (in the selected region)} videos. \\
\textbf{Step 3:} When selecting a video to watch, the user is redirected to a page as shown in Fig.~\ref{fig:exp-screenshot}. 
   $(a)$ The user watches the video
   . Some videos may contain rebufferings, which are artificially added to emulate a low QoS environment.
   $(b)$ $5$ videos are recommended to the user to watch next. 
    $(c)$ The user is requested to rate her viewing experience by rating 4 parameters: \textit{interest} in the content of the video (\textit{Int}), satisfaction from the video quality (\textit{QoS}), relevance of the recommendations (\textit{QoR}), and overall enjoyment (\textit{QoE})
    .\\
\textbf{Step 4:} Each time the user selects a video from the recommendations, step 3 is repeated. The maximum number of videos to watch is $5$. After the fifth video, the experiment session ends. 

\myitem{}\indent The information communicated to the users (when they enter the platform) is that they are going to select, watch, and rate a series of five YouTube videos, to some of which we may artificially introduce rebuffering for the purposes of our research study. No further information is revealed about how we recommend videos and the QoS impairments, to avoid biasing their selections and ratings. We also inform the users that no personal information is collected.

\removeTextForCARSsubmission{
\begin{figure}
\centering
\includegraphics[width=0.4\linewidth]{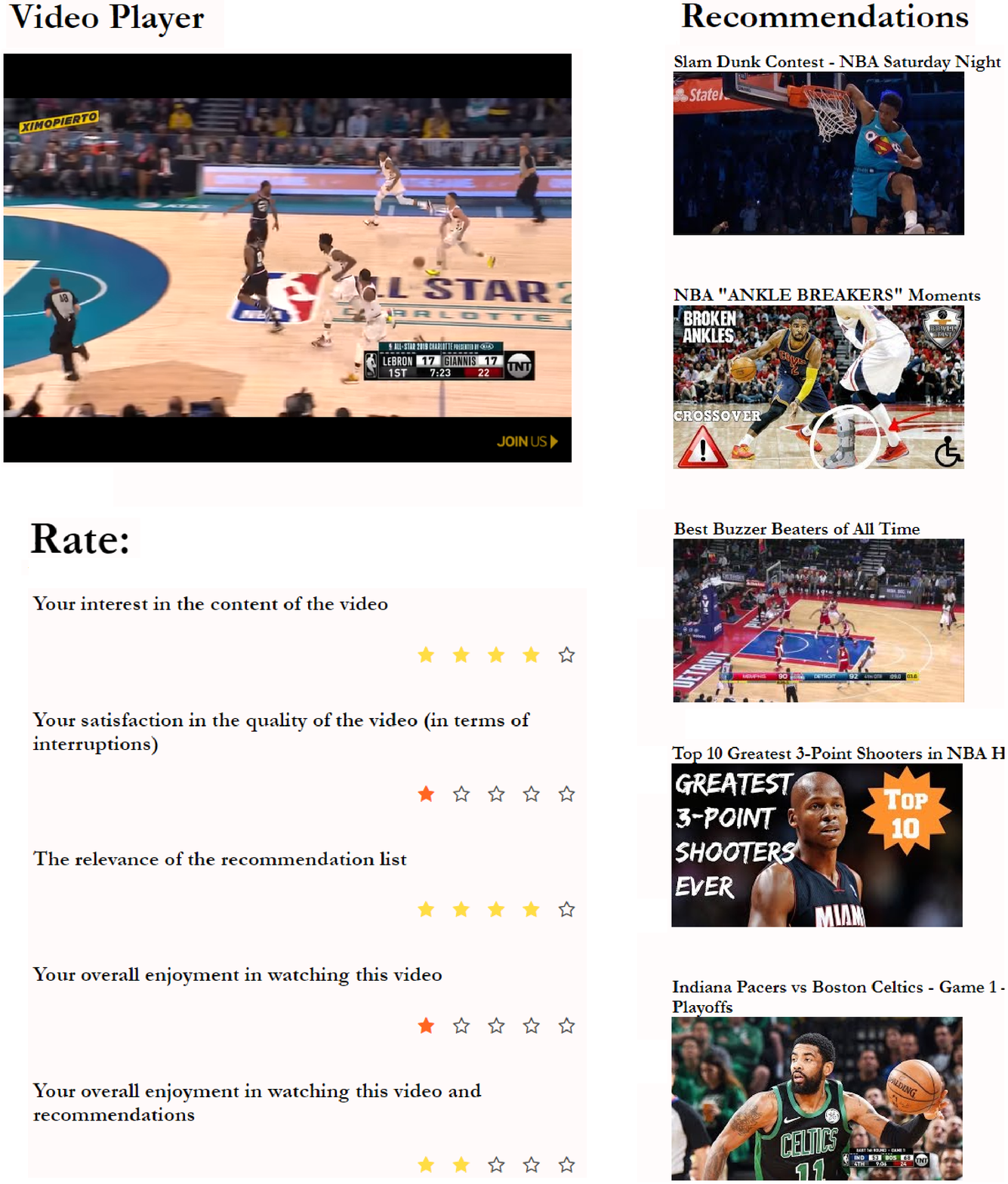}
\caption{Experimental platform - instance of a user experiment: (i) a user watches a video (top/left), and is requested to (ii) rate her satisfaction from the watched video and recommendations (bottom/left) and (iii) select one of the recommendations to proceed to the following video (right).}
\label{fig:exp-screenshot}
\end{figure}
}

\begin{figure}
\begin{minipage}{\mycustomwidth}
\centering
\includegraphics[width=\linewidth]{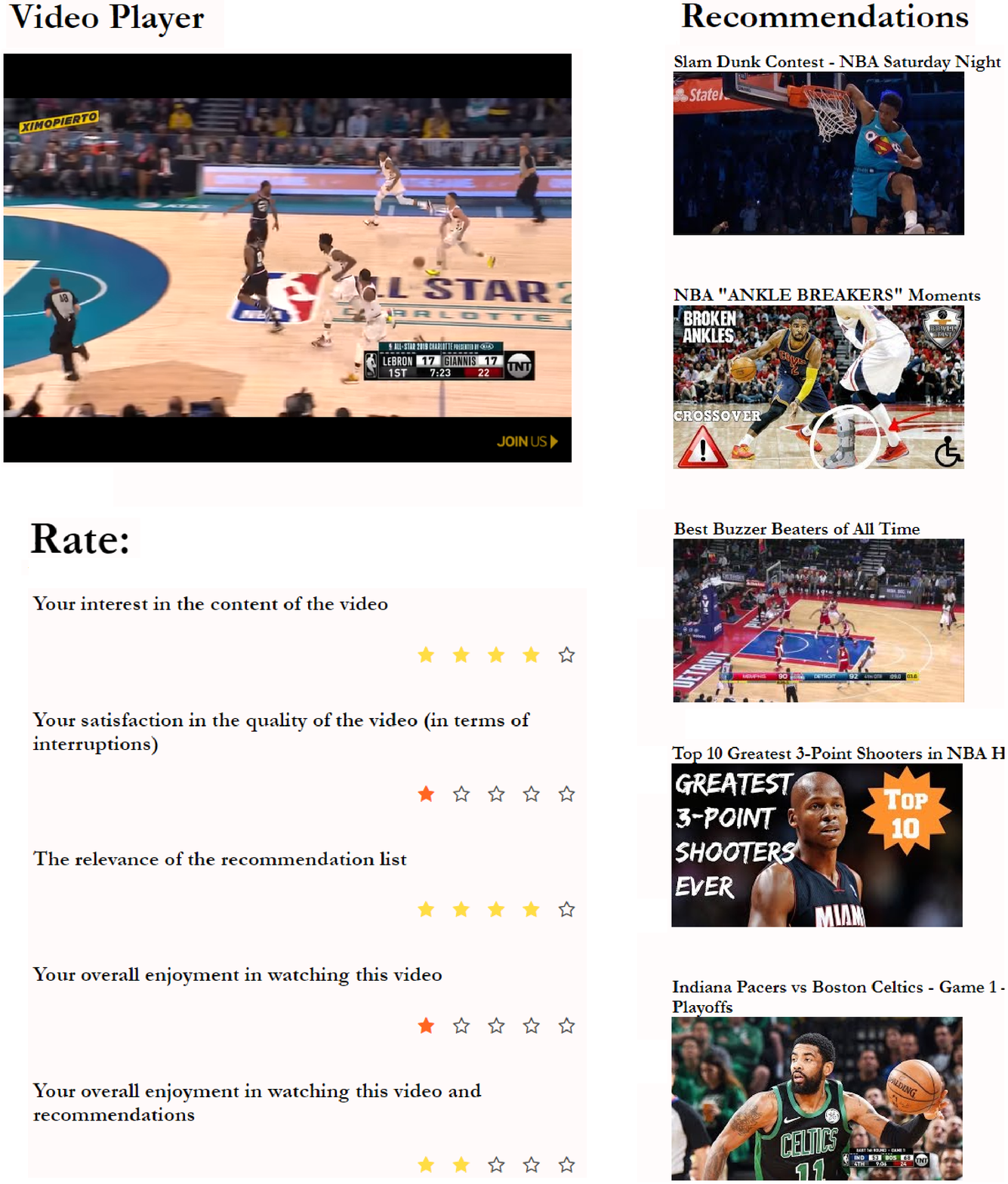}
\caption{Experimental platform - instance of a user experiment: (i) a user watches a video (top/left), and is requested to (ii) rate her satisfaction from the watched video and recommendations (bottom/left) and (iii) select one of the recommendations to proceed to the following video (right).}
\label{fig:exp-screenshot}
\end{minipage}
\hfill
\begin{minipage}{\mycustomwidth}
\begin{minipage}{\linewidth}
\centering
\includegraphics[width=\mycustomcolumnwidth]{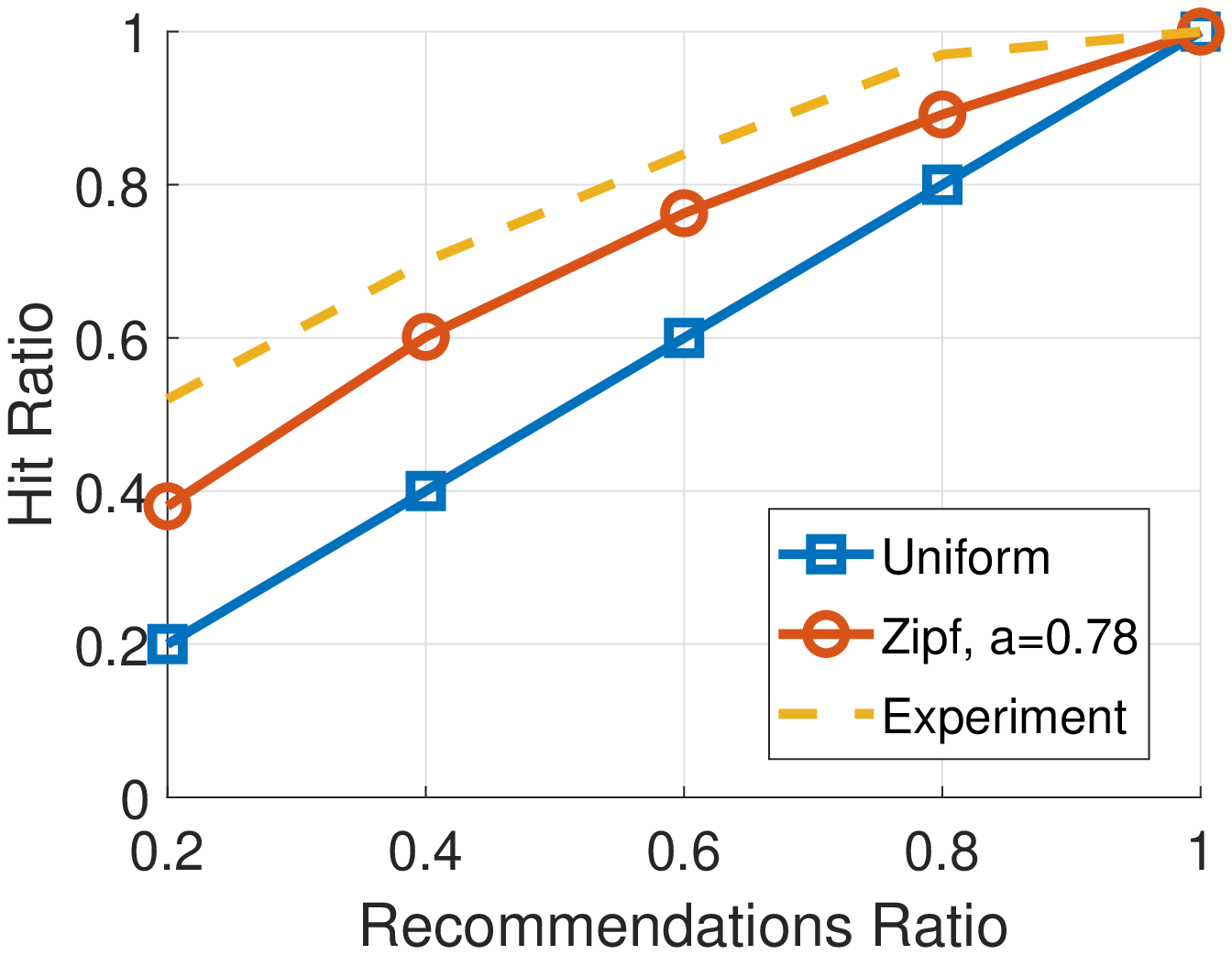}
\caption{\textit{HR} as a function of \textit{RR}.}
\label{fig:chr-crr}
\end{minipage}
\vspace{0.75cm}
\\
\begin{minipage}{\linewidth}
\centering
\includegraphics[width=\mycustomcolumnwidthA]{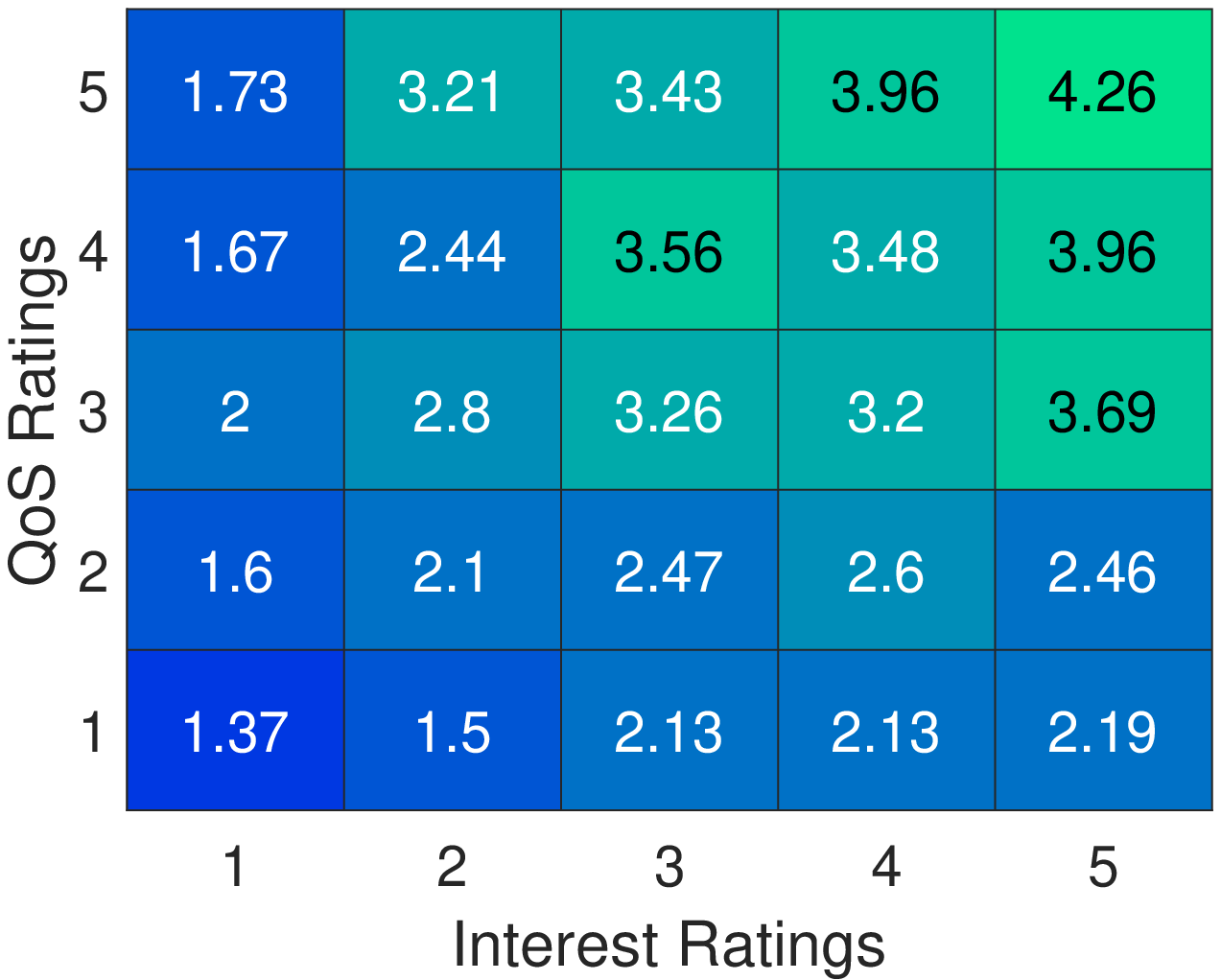}
\caption{\textit{QoE} as a function of \textit{QoS} and user \textit{interest}. }
\label{fig:qoe-of-int-qos}
\end{minipage}
\end{minipage}
\end{figure}

\subsection{Experiment Setup}

\myitem{Region (Step 1).} We offer as options a subset of the regions provided by the YouTube API~\cite{youtube-api}; we selected $7$ representative regions (different continents, diverse demographics, available video data). 

\myitem{Initial list of videos (Step 2).} For each region, we retrieve from the YouTube API the list of $50$ top trending videos. We randomly select $20$ of them (for the selected region) to present to the user.

\myitem{QoS impairments (Step 3a).} We assumed that the network could deliver a set of videos in high QoS, and all the remaining in lower QoS. For instance, this may correspond to the case of a congested network, where some contents are stored in a cache close to the user, thus not affected by congestion. The remaining are assumed to be fetched from remote servers, with congestion-imposed impairments. As an initial effort to emulate such scenarios, we assumed for this experiment that the available bandwidth for low QoS videos is lower than the video bit-rate, and emulated 1sec. of rebuffering every 8sec. of the video. 
\bb{Rebufferings are one of the most evident type of QoS impairments}
~\cite{nam2016qoe,doan2019tracing}; considering more types of QoS impairments (e.g., bit-rate changes, start-up delays) could be a part of future work.

\emph{Choice of high-QoS videos:} For the purposes of this experiment, we assumed a list of $500$ videos can be delivered in high QoS (e.g., assumed to be cached)
; we consider a different list per region. In each list, we select to first include the top $50$ trending videos in this region. Then, for each of these $50$ videos, we request its $50$ recommendations / related videos provided by YouTube API. From these $2500$ ($=50\cdot50$) videos, we add in the list the $450$ videos with the higher number of views (``most popular''). We stress that this would not be necessarily the ``optimal'' caching policy (e.g., see~\cite{sermpezis2018soft,chatzieleftheriou2019jointly,zhu2018coded}), but rather a first-cut effort to differentiate between high and low QoS files. Using a more sophisticated caching policy is orthogonal to the findings of this work. 

\myitem{List of recommendations (Step 3b).} The list of the $5$ recommendations given to the user when watching a video are either (i) the original YouTube recommendations (i.e., the top 5 related videos results returned from the YouTube API), or (ii) derived from a QoS-aware recommendation algorithm. Regarding the latter option,  various algorithms have been recently proposed~\cite{chatzieleftheriou2019jointly, kastanakis2018cabaret, giannakas2018show}, of varying complexity. We choose to use the QoS-aware algorithm of~\cite{kastanakis2018cabaret}, due to its simplicity and relevance to YouTube.\footnote{Since the goal of this paper is not to optimize a potential QoS-aware RS algorithm, but simply to investigate the effect of QoS-aware recommendation nudging, we stress that the analysis to follow is applicable to any other QoS-aware algorithm.} The algorithm of~\cite{kastanakis2018cabaret} does a breadth-first search (up to depth 2) in the related videos provided from the YouTube API. If the algorithm finds among these videos at least 5 high-QoS related files (we pre-assign a subset of files as, for example, cached) it recommends them; if the search finds less than 5 related videos to be high-QoS, it completes the list of recommendations with the original top YouTube recommendations.

\myitem{Collected data (Step 3c).}
In each experiment session we collect the following data: %
(i) ID of watched video, (ii) ID of videos in the final recommendation list (i.e., the $5$ videos in the right side in Fig.~\ref{fig:exp-screenshot}), and their positions in this list, (iii) ID of videos in the initial list of recommendations produced by YouTube (this list is not presented to the user), (iv) ID of videos available in high-QoS, (v) ratings of the user for \textit{Interest}, \textit{QoS}, \textit{QoR}, \textit{QoE}\footnote{Note that the \textit{Interest}, \textit{QoS} and \textit{QoE} are ratings/metrics that refer to the watched video, while the \textit{QoR} refers to the videos in the recommendation list.}.



\section{Results}\label{sec:results}
We conducted an experimental campaign \myblue{recruiting participants through mailing lists and social media. We }
collected 742 samples 
from users in \bb{ 
North/South America, Europe and Asia
}. In the following we present main findings from the analysis of the results. Our aim is to provide initial insights for the applicability of QoS-aware RSs (from the user and content provider perspective), the involved trade-offs, and directions that require further investigation.


\myitem{\textit{Finding 1: Nudging recommendations towards high-QoS videos is not perceived as intrusive.}} 

One important concern is that users might perceive the ``nudged'' recommendations as significantly ``worse'', and not be willing to select them. This could nullify the expected impact on both network performance and QoS-aware recommendations. 
For this reason, we first study whether users are willing to select the ``nudged'' recommendations. To quantify this, we define two metrics: (i) the \textit{hit ratio}, \textit{HR}: this is the ratio of high-QoS videos that users \emph{selected}, over the total number of viewed videos, i.e., the ``click-through'' rate for the nudged videos; (ii) the \textit{recommendation ratio}, \textit{RR}: this is the ratio of recommended high-QoS videos over the total number of recommended videos. 

In order to better understand how these metrics attempt to capture the impact of ``nudging'', consider the following: Say 2 out of 5 recommendations were ``nudged'', i.e., 2 of the original 5 recommendations are now replaced with 2 high QoS videos (according to the scheme described earlier). If the user picked uniformly among the original 5 recommendations, then nudging could be considered non-intrusive if the click-through rate for the 2 high QoS videos remains around $40\%$ (2 out of 5), while lower values would suggest that users tend to disregard the nudged recommendations.

\removeTextForCARSsubmission{
\begin{figure}
\centering
\begin{minipage}[t]{\mycustomwidth}
\centering
\includegraphics[width=\mycustomcolumnwidth]{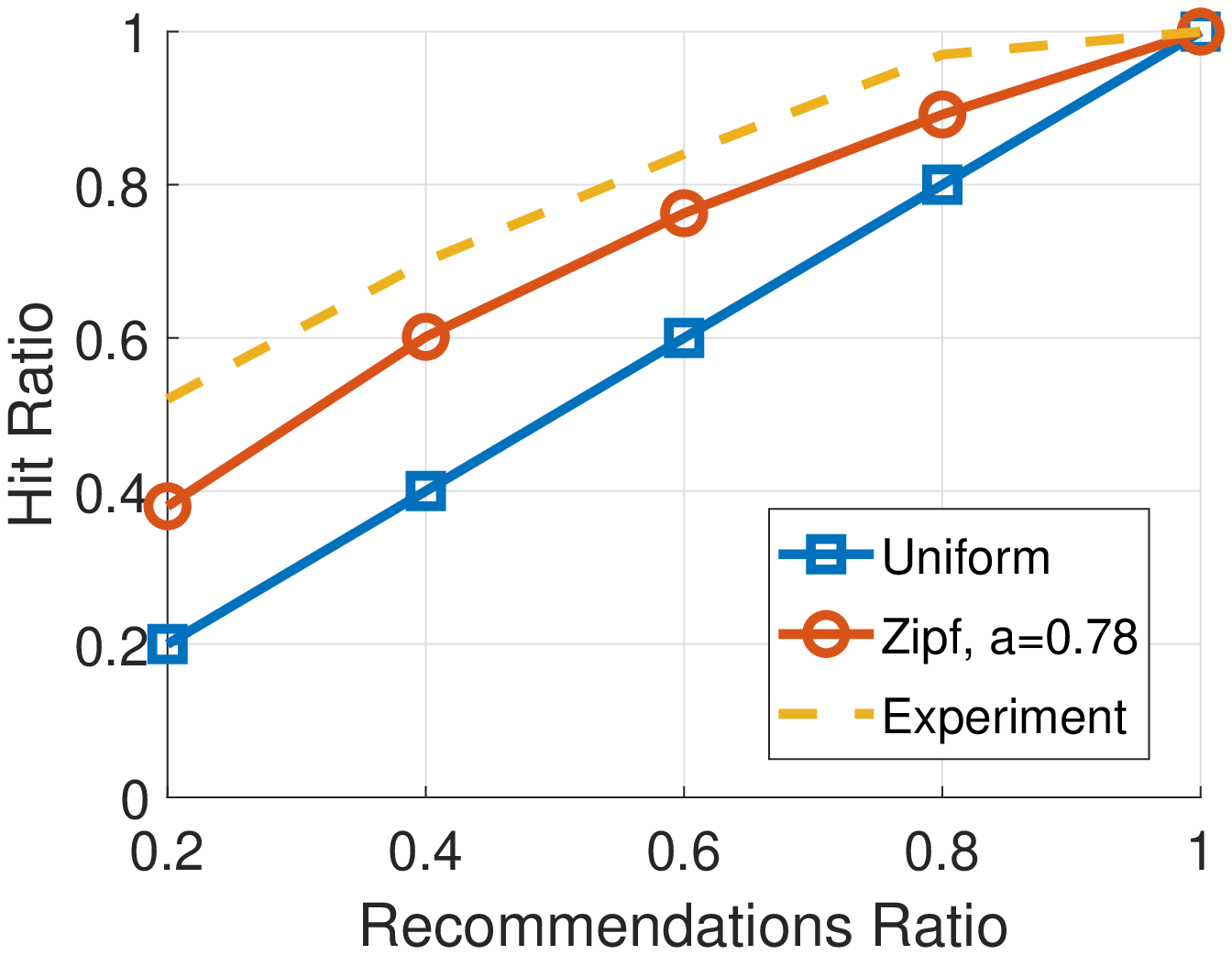}
\caption{\textit{HR} as a function of \textit{RR}.}
\label{fig:chr-crr}
\end{minipage}
\hspace{0.05\linewidth}
\begin{minipage}[t]{\mycustomwidth}
\centering
\includegraphics[width=\mycustomcolumnwidthA]{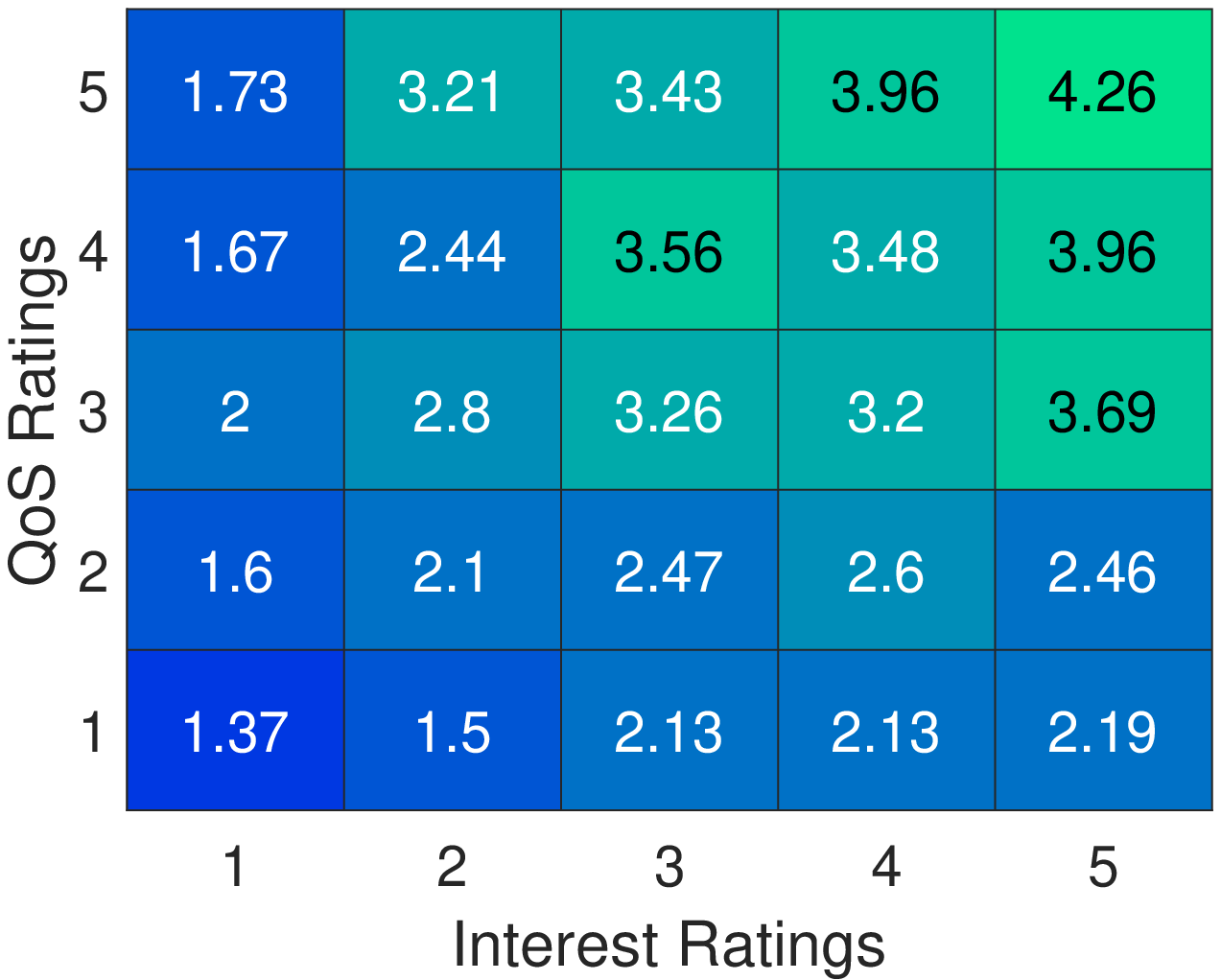}
\caption{\textit{QoE} as a function of \textit{QoS} and user \textit{interest}. }
\label{fig:qoe-of-int-qos}
\end{minipage}
\end{figure}
}

Figure~\ref{fig:chr-crr} investigates this behavior. We first plot the \textit{HR} (y-axis) as a function of \textit{RR} (x-axis) that would be achieved if users selected each time randomly one of the 5 recommended videos, as explained earlier (blue line - ``uniform''). We also plot the respective HR values observed in our experiments (yellow dashed line). For example, this plot shows that among all the sessions where only 2 out of 5 recommendations were for high-QoS videos ($x=2$), the users selected one of these 2 recommendations in almost $70\%$ of the cases ($y\approx 70\%$).

Hence, the \textit{HR} (click-through rate) for nudged videos is not just close to the uniform average (the desired behavior), but well above, regardless of the \textit{RR} value (i.e., how many of the recommendations are nudged). While this might be surprising, we conjecture that the extra ``gain'' in HR comes from the fact that we were placing the high QoS video(s) on the top of the recommendation list: 
users tend to favor content higher up the list, even if videos are equally interesting~\cite{RecImpact-IMC10,sapiezynski2019quantifying}. Hence, for a fairer comparison, we also plot the expected click-rate taking into account content position (red line - ``Zipf'')\footnote{The probability $p_{i}$ to select the content at the $i^{th}$ position ($i=1,...,5$) follows a Zipf law $p_{i}\sim\frac{1}{i^{a}}$ with $a=0.78$~\cite{RecImpact-IMC10}}. The experimental results are still close to this more realistic YouTube-like behavior (``Zipf'').



Summarizing, the above observations provide useful evidence that \textit{using a carefully designed QoS-aware RSs would not have a negative impact on user preferences and/or the performance of the RS}. This finding is particularly relevant  
in light of recent results indicating that QoS-aware RSs  bring significant gains to the network performance~\cite{sermpezis2018soft,chatzieleftheriou2019jointly,kastanakis2018cabaret,zhu2018coded,lin2018joint,giannakas2018show,song2018making}.   \textit{We are the first to provide experimental evidence for the feasibility (in terms of user perception/acceptance) of QoS-aware~RSs.}

\myitem{\textit{Finding 2: QoS-aware recommendations bring a positive impact on the overall user satisfaction
.}}

A user would enjoy more a high-QoS video, among two videos in which she is equally interested. However, nudging recommendations implies that we sometimes offer less interesting videos, with the intent to make up for this with better QoS. Yet, what is the impact of this action on the user satisfaction? Will the user enjoy more a ``less interesting'' video? 

Table~\ref{table:user-ratings} shows the average user ratings for the \textit{QoS}, \textit{interest}, \textit{QoR}, and \textit{QoE} among all high-QoS and low-QoS videos\footnote{\myblue{For the QoR, we consider the ratings at the previous session of a video viewing, i.e., the quality of the recommendation list that included the viewed video; the other ratings are from the video session itself.}}. We see that the impact on user perception of recommendation quality (\textit{QoR}) is not significant for the nudged recommendations. Moreover, the difference on users \textit{interest} in the viewed videos is negligible. This indicates that \textit{nudging recommendations towards high-QoS content does not affect negatively the user interest for the selected content}. 
\begin{table}[h]
\caption{Average ratings \myblue{($\pm$ confidence intervals), in 1 to 5 stars $\bigstar$, for the groups of Low-QoS and High-QoS videos}
}
\label{table:user-ratings}
\begin{tabular}{c|cccc}

{} & Low-QoS Videos & High-QoS Videos\\
\hline
{QoS}  &  1.87 ($\pm$0.14) & 4.30 ($\pm$0.11)\\
{Interest} &  3.54 ($\pm$0.16)  & 3.61 ($\pm$0.14)\\
{QoR} & 3.60 ($\pm$0.15) & 3.40 ($\pm$0.13)\\
{QoE} & 2.40 ($\pm$0.20) & 3.67 ($\pm$0.16)
\end{tabular}
\end{table}

While this is definitely promising, the more important observation is that \textit{providing QoS-aware recommendations has a very positive effect on the total user satisfaction}. The results in Table~\ref{table:user-ratings} (i) verify our intuition (and motivation of our work) that the impact on \textit{QoS} is clearly perceived by the users (cf. the large difference in average ratings for \textit{QoS}), and (ii) demonstrate that the \textit{QoS is a main factor affecting the overall user experience (QoE)}, i.e., a low-QoS video streaming leads to poor user experience \textit{QoE}. \myblue{In fact, applying a chi-squared test in our data, the null hypothesis ``QoS and QoE are independent'' is rejected with certainty (i.e., with a p-value $\approx10^{-54}$).}



Figure~\ref{fig:qoe-of-int-qos} presents in more detail the joint impact of \textit{interest} and \textit{QoS} on the user experience (\textit{QoE}). A video available in low-QoS (e.g., for 1 or 2 in the y-axis) leads always to poor user experience (less than 3 stars), even when the users are very interested in the content of the video (e.g., 5 in x-axis). On the contrary, we observe that even when the users are not very interested in the video, their experience can be moderately good when the video is provided in high-QoS; cf., for example, the average $QoE$ is higher than 3 for 2 in the x-axis (Interest) and 5 in the y-axis (QoS). 

In general, the factors \textit{QoS} and user \textit{interest} affect very differently and in a complex (non-linear) way the QoE. For example, denoting $\{x,y\}$ the values for \textit{interest} and \textit{QoS}, we observe that while $\{2,5\}$ gives a higher \textit{QoE} than $\{5,2\}$, an opposite trend appears between the pairs $\{3,5\}$ and $\{5,3\}$. 
We believe that the findings in Fig.~\ref{fig:qoe-of-int-qos} motivate the need for a detailed investigation of this interplay, which could further lead to better RS design\bb{; Section~\ref{sec:ml-model} is towards this direction}.

\myitem{\textit{Finding 3: QoS is a factor of (at least) equal significance to user interest for the retention rate in video services.}}

The results above indicate that QoS-aware recommendations can benefit the users. But, is it worthwhile for content providers to adjust their RS to take QoS into account? Could this benefit their services, e.g., the retention rates, as well?

\begin{figure}
\centering
\begin{minipage}[t]{\mycustomwidth}
\centering
\includegraphics[width=\mycustomcolumnwidth]{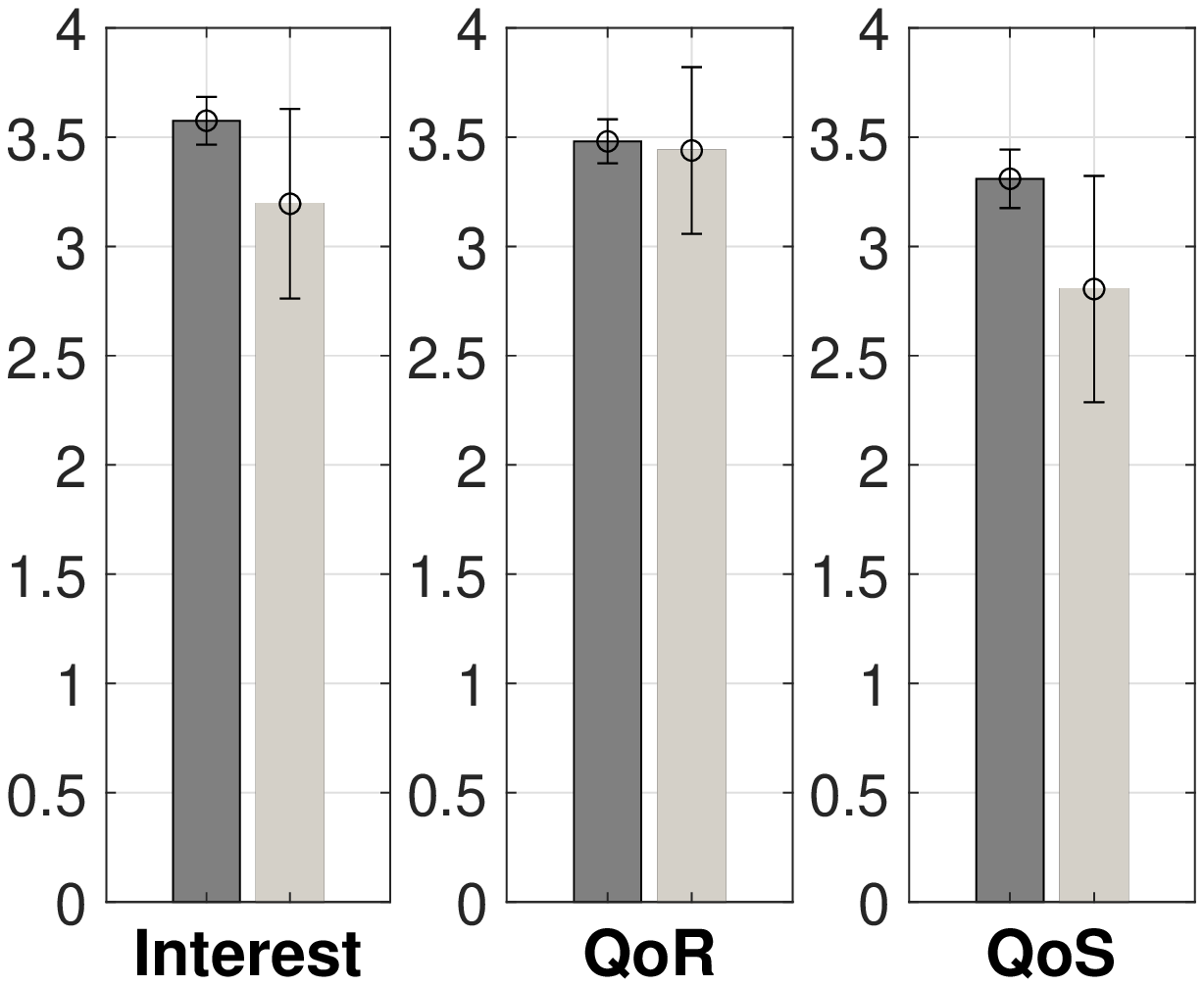}
\caption{Avg. Rating in \textit{all} (dark bars) vs. \textit{abandoned} (light bars) sessions.}
\label{fig:abandonmentAvg}
\end{minipage}
\hspace{0.05\linewidth}
\begin{minipage}[t]{\mycustomwidth}
\centering
\includegraphics[width=\mycustomcolumnwidth, trim={0.7cm 0 0.75cm 0},clip]{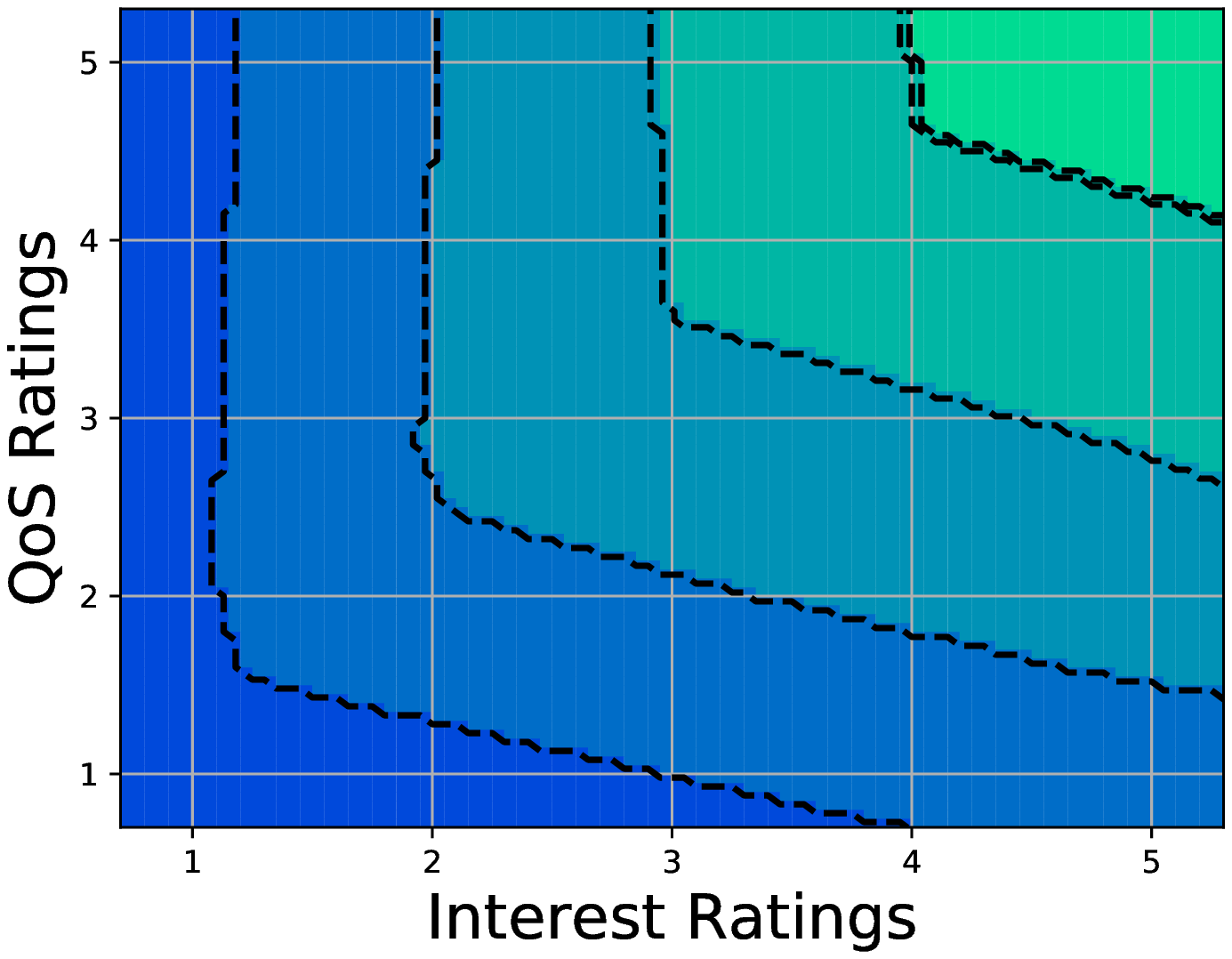}
\caption{MLP regression model for \textit{QoE} with features \textit{QoS} and user \textit{interest}.}
\label{fig:2D-qos-int-MLPR}
\end{minipage}

\end{figure}

We provide an initial answer to this in Fig.~\ref{fig:abandonmentAvg}, where we present the average ratings (y-axis) for \textit{interest}, \textit{QoS}, and \textit{QoR} (x-axis), for \textit{all video sessions} (dark color bars) and for \textit{sessions after which the users abandoned the experiment} (light color bars)\footnote{The last video sessions (after which the experiment ends) are not considered as abandonment and thus are not taken into account in these calculations.}. These results can be interpreted as an analogy to the reasons that make users to leave the video service. 
As we can see, the difference in the average \textit{QoR} ratings is negligible, while the ratings for \textit{interest} and \textit{QoS} are around $10\%-15\%$ lower in the abandonment sessions (the t-test rejects the null hypothesis of \textit{interest}/\textit{QoS} having equal mean in all and abandonment sessions with a 5\% confidence level). This indicates that low interest in the content of the video or low quality of the video tend to affect the decision of users to continue watching videos. While this is intuitive, our results quantify the effect of each factor. Moreover, the no-impact of QoR implies that even if good recommendations are provided to a user for future sessions, she may still leave the service due to poor experience.

\section{Modeling User Experience}
\label{sec:ml-model}
The analysis of our results, e.g., see Fig.~\ref{fig:qoe-of-int-qos}, indicate that non-linear trends may exist in the relation between user \textit{interest}, \textit{QoS}, and \textit{QoE}. 
This highlights the need to go beyond previously proposed simple linear models~\cite{ding2014combining} which may not be adequate in practice. Hence, our goal in this section is to better understand and model how the \textit{interest} and \textit{QoS} jointly affect the experience of a user.

\myitem{Overview of modeling approach.} Towards achieving the above goal, we select to model the interplay of \textit{QoS} and \textit{interest} with generalized linear models (GLM), which allow to capture complex effects and at the same time are easily interpretable and can reveal insights.
\bb{This choice is in line with the increasing demand (in the AI community in general~\cite{weld2019challenge}, and the RS community in particular~\cite{dacrema2019we}) for explainable ML models and AI algorithms.} For instance, building and training a deep learning model that could fit our data (which are collected from an experimental testbed), but could not be interpreted, would be of limited practical value (e.g., for fitting data from an operating video service).

In the following, we first carefully select the set of features to be used in such a GLM (Section~\ref{sec:feature-selection}), and then we design a model to predict QoE and analyze its structure to derive interesting insights (Section~\ref{sec:model-and-insights}).
The results show that our approach can achieve a good trade-off between simplicity/explainability and accuracy.

\subsection{Feature Selection}\label{sec:feature-selection}

\myitem{Basic features
.} The basic features that affect the user experience, and for which we have explicit user ratings, are the \textit{QoS} and \textit{interest}. While we have explicit ratings for \textit{QoR} as well, this factor does not significantly affect the user enjoyment in the current video\footnote{We   verified this by considering the \textit{QoR} ratings in the models we present in this section, and observing non-significant differences.
}. In the remainder, we denote the basic features as $x_{1}=QoS$ and $x_{2}=Int$.

\myitem{Meta-features.} To capture the more complex effects that the basic features have on the \textit{QoE}, we can use ``meta-features'', namely, (i) non-linear functions of single features $g(x_{i})$ (e.g., squares, logarithms), and/or (ii) complex functions of combinations of features $g(x_{1},x_{2})$ (e.g., products, ratios)\bb{, similarly to common intelligible ML practices~\cite{lou2012intelligible-gam,caruana2015intelligible-ga2m}}. However, there are unlimited options for possible meta-features, and we need to narrow down the set of the meta-features we consider and select only the most important. 

\textit{(i) Univariate meta-features $g(x_{i})$:} We start from the single feature functions $g(x_{i})$ and test a wide range of functions $g$ as predictors of the QoE ratings, i.e., $\hat{QoE}=g(x_{i})$. We find that linear functions, $g(x_{i})=\alpha\cdot x_{i}+\beta$, perform comparably to non-linear functions. Thus, including non-linear \textit{univariate} functions $g(x_{i})$ as meta-features would not significantly improve a model's predictive power.

\textit{(ii) Multivariate meta-features $g(x_{1}, x_{2})$:} We proceed to combinations of feature pairs, and test different 
regression models/functions $\hat{QoE}=g(x_{1},x_{2})$. Table~\ref{tab:ml-model-fit-mae} presents the prediction performance 
of the models we tested: specifically, we first removed ``outliers'' from our dataset (i.e., samples where the \textit{QoE} rating is higher/lower than both \textit{QoS} and \textit{interest}; to avoid cases where users misinterpreted the quantities they were asked to rate, and obtain clearer insights), \bb{and used 5-fold cross validation to train the models and calculate the mean absolute error (MAE)}. 
%
The results show that the Multi-layer perceptron (MLP) regression model, in which we used a neural network with two hidden layers with 16 nodes each, achieves the lowest MAE and thus provides the best meta-feature $g(x_{1},x_{2})$.
However, MLP does not give an (interpretable) closed-form function $g(x_{1},x_{2})$, which would be desired. Hence, we use a visualization of its predictions in Fig.~\ref{fig:2D-qos-int-MLPR} (axes correspond to the feature values, from 1 to 5, and colored areas denote the QoE predictions: dark blue corresponds to 1 and light green to 5
). A careful inspection of the visualization reveals an interesting pattern: the MLP meta-feature can be approximated well by a function $g(QoS,Int)\propto \min\{QoS, Int\}$
\footnote{An equivalent approximation could be $g(QoS,Int)\propto QoS \cdot Int$; in our dataset the quantities $\min\{QoS, Int\}$ and $QoS \cdot Int$ have a correlation coefficient 0.98.}.


\textit{Remark:} For completeness, we tested several other functions $g(x_{1},x_{2})$ as well (such as ratios, maximum, etc.), which were found to not be statistically significant for the QoE predictions
.

\myitem{Summarizing, the selected features are:} $x_{1}=QoS,~\text{     }~x_{2}=Int,~\text{     }~x_{3} = min\{QoS,Int\}$


\removeTextForCARSsubmission{
\begin{table}[h]
\centering
\caption{Mean absolute error, MAE, of different regression models $\hat{QoE}=g(QoS,Int)$.}
\label{tab:ml-model-fit-mae}
\vspace{-0.5\baselineskip}
\begin{tabular}{l|c||l|c}
{Models (linear)} &  {MAE} & {Models (non-linear)} &{MAE}\\
\hline
{Linear Regression} &  {0.56} &  {Decision Tree Regr.}    &  {0.46}\\
{Logistic Regression}  &  {0.50} &  {SVR}       &  {0.46} \\
{Ordinal Regression}       &  {{0.48}} &  {MLP Regression}       &  {\textbf{0.44}}
\end{tabular}
\end{table}
}

\begin{table}
\begin{minipage}[t]{\mycustomwidth}
\centering
\caption{Mean absolute error, MAE, of different regression models $\hat{QoE}=g(QoS,Int)$.}
\label{tab:ml-model-fit-mae}
\vspace{-0.5\baselineskip}
\begin{tabular}{l|c}
\hline
{Models (linear)} &  {MAE}\\
\hline
{Linear Regression} &  {0.56} \\
{Logistic Regression}  &  {0.50} \\
{Ordinal Regression}       &  {{0.48}}\\
\hline
\multicolumn{2}{c}{} \\
\hline
{Models (non-linear)} &{MAE}\\ 
\hline
{Decision Tree Regr.}    &  {0.46}\\
{SVR}       &  {0.46} \\
{MLP Regression}       &  {\textbf{0.44}} \\
\hline
\end{tabular}
\end{minipage}
\hfill
\begin{minipage}[t]{\mycustomwidth}
\centering
\caption{Mean absolute error (MAE) and predictions accuracy of three GLMs and two baseline models.}
\label{tab:performance-ordinal}
\vspace{-0.5\baselineskip}
\begin{tabular}{c|c|c|ccc}
{} &Model & MAE & \multicolumn{3}{c}{prediction error}   \\
{} &{} & {} & \multicolumn{3}{c}{$|\hat{QoE}-QoE|$:}   \\
{}& {} & {} & {$=0$}& {$=1$}& {$>1$}\\
\hline
\multirow{3}{*}{\rotatebox[origin=c]{90}{GLMs}}
    & Linear & {0.47} & {60\%} & {37\%} & {3\%}\\
    & Logistic & {0.42} & {66\%} & {29\%} & {5\%}\\ 
    & Ordinal & {\textbf{0.41}} & {62\%} & {33\%} & {5\%}\\
\hline
\multirow{3}{*}{\rotatebox[origin=c]{90}{}} 
    &Dummy & {1.25} & {19\%} & {36\%} & {45\%}\\
    & Vanilla-RS & {0.80} & {49\%} & {33\%} & {18\%}\\
\end{tabular}
\end{minipage}
\end{table}

\subsection{Model and Insights}\label{sec:model-and-insights}

We proceed to design an interpretable model, namely, a GLM that uses a linear combination of the selected features, i.e., $\hat{QoE} = f(\sum_{i} w_{i}\cdot x_{i})$, where $w_{i}$ is the weight of the feature $x_{i}$, and $f$ the model function. 
We consider three different GLMs, namely, \textit{Linear}, \textit{Logistic}, and \textit{Ordinal} regression\footnote{The model functions for Linear and Logistic Regression are $f(x)=x$  and $f(x)= \frac{1}{1+e^{-x}}$, respectively. Ordinal Regression is a variant of Logistic Regression, applied in cases where the variables take values in an ordered scale (e.g., 1 to 5), and is common for modeling human levels of preference.}
. We train the models, and calculate their MAE and accuracy.


\myitem{\textit{The Ordinal Regression model predicts in 95\% of the cases the QoE within $\pm1$ of the real user rating.}} %
Table~\ref{tab:performance-ordinal} shows that the Ordinal Regression is the best performing model (Logistic Regression has comparable accuracy): the \textit{QoE} predictions deviate on average from the real ratings by $\pm 0.41$ (in a 1-5 scale). A more detailed look shows that 62\% of the predictions are exact, and 95\% are within $\pm1$ of the real rating; only 5\% predictions clearly cannot capture the experience declared by the user.

\removeTextForCARSsubmission{
\begin{table}[h]
\centering
\caption{Mean absolute error (MAE) and predictions accuracy of three GLMs and two baseline models.}
\label{tab:performance-ordinal}
\vspace{-0.5\baselineskip}
\begin{tabular}{c|c|c|ccc}
{} &Model & MAE & \multicolumn{3}{c}{prediction error, $|\hat{QoE}-QoE|$:}   \\
{}& {} & {} & {$=0$}& {$=1$}& {$>1$}\\
\hline
\multirow{3}{*}{\rotatebox[origin=c]{90}{GLMs}}
    & Linear & {0.47} & {60\%} & {37\%} & {3\%}\\
    & Logistic & {0.42} & {66\%} & {29\%} & {5\%}\\ 
    & Ordinal & {\textbf{0.41}} & {62\%} & {33\%} & {5\%}\\
\hline
\multirow{3}{*}{\rotatebox[origin=c]{90}{}} 
    &Dummy & {1.25} & {19\%} & {36\%} & {45\%}\\
    & Vanilla-RS & {0.80} & {49\%} & {33\%} & {18\%}\\
\end{tabular}
\end{table}
}

\myitem{\textit{Taking QoS into account reduces more than 3 times the cases where the QoE predictions fail significantly.}} To better quantify the added value of the proposed model, we compare it with two baseline models, namely: (i) \textit{Dummy}: a model used for sanity check~\cite{sk-learn-dummy,scikit-learn} that always predicts $\hat{QoE}=3$; (ii) \textit{Vanilla-RS}: a RS that does not take QoS into account and predicts $\hat{QoE}=Int$ (similarly to today's RSs). The results of Table~\ref{tab:performance-ordinal} show that the proposed model performs 3 times better than the worst-case (\textit{Dummy}), and achieves 2 times lower MAE than a QoS-unaware RS (\textit{Vanilla-RS}). The \textit{Vanilla-RS} fails significantly (error $>\pm1$) to predict the user experience 3.6 more times than the proposed model (18\% vs. 5\%). Finally, the importance of using the meta-feature $\min\{QoS,Int\}$ can be seen by comparing to the MAE of the corresponding model in Table~\ref{tab:ml-model-fit-mae}, whose MAE (0.48) is 15\% larger.

\begin{figure}
\centering
\begin{minipage}{\mycustomwidth}
\centering
\begin{tabular}{c|c}
Feature     &  Weight\\
 $x_{i}$ &  $w_{i}$\\
\hline
$QoS$ & {0.17} \\
$Int$ & {0.30} \\
$\min\{QoS,Int\}$ & {0.53}\\
\end{tabular}
\end{minipage}
\hspace{0.05\linewidth}
\begin{minipage}{\mycustomwidth}
\centering
\includegraphics[width=\mycustomcolumnwidthB]{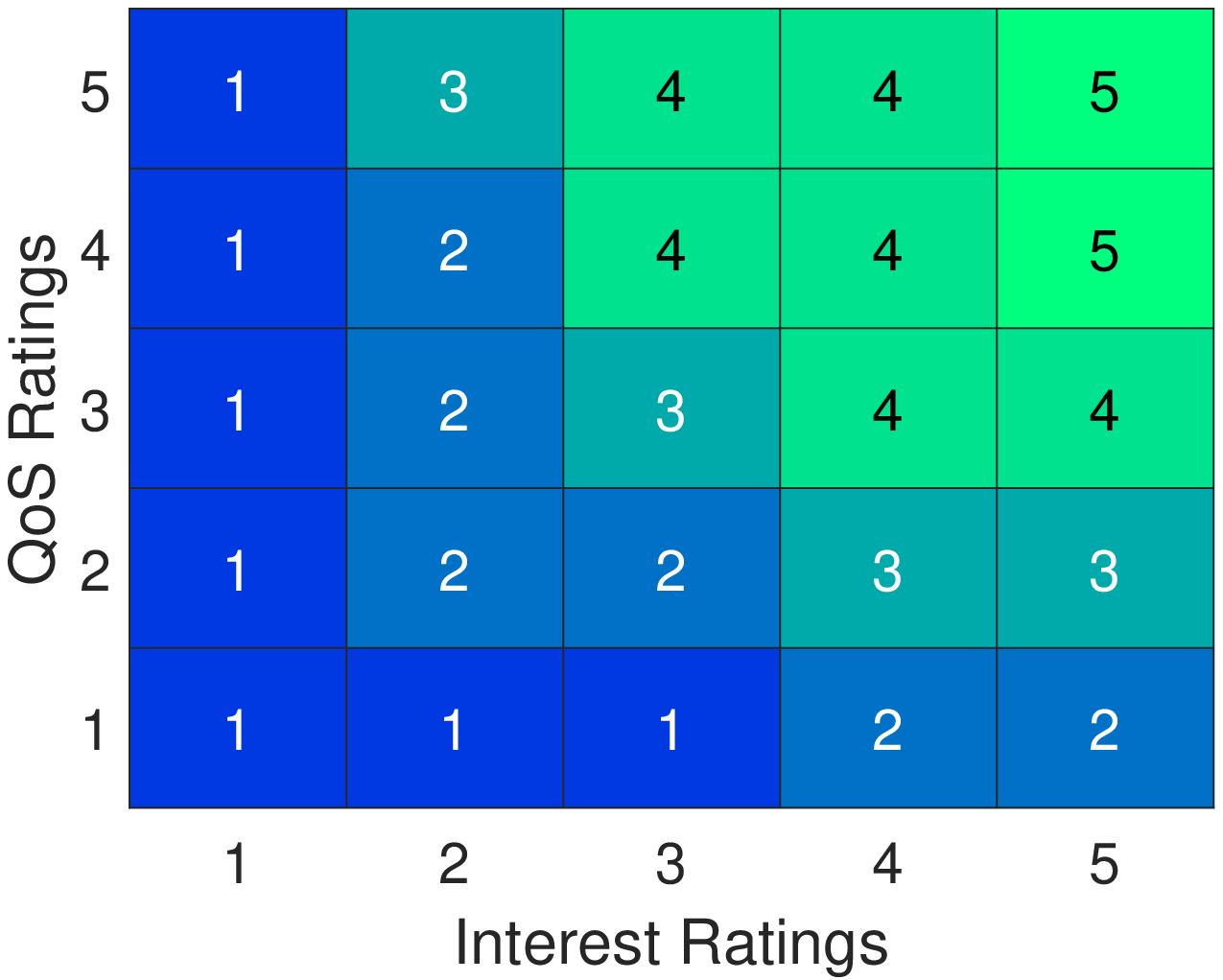}
\end{minipage}
\caption{Parameters (left) and QoE predictions (right) of the proposed Ordinal Regression model
.}
\label{fig:model-heat}
\end{figure}

\myitem{\textit{The interpretation of the model suggests that a RS should get both \textit{QoS} and user \textit{interest} to satisfy a minimum standard; then the RS can pay more attention to user \textit{interest}.}} The main motivation for using a GLM is interpretability of results. In particular, the weights $w_{i}$ indicate the relevant importance of the feature $x_{i}$, i.e., the larger the absolute value of the weight, the higher the importance of the meta-feature in the model. In Fig.~\ref{fig:model-heat} (left) we present the weights of the features in the proposed model.
%
A main observation is that the minimum of the user \textit{interest} and \textit{QoS} plays the most significant role. This clearly indicates that \textit{users cannot be highly satisfied unless both the video is delivered in high-QoS and lies within their interests}; in other words, recommending a video with poor QoS is equally bad to a non-relevant recommendation. This finding further supports the motivation of our paper, and stresses the need for QoS-aware RSs in multimedia services. 

Figure~\ref{fig:model-heat} (right) visualizes the proposed model, and allows to directly compare \{QoS,Int\} options for potential recommendations; e.g., \{4,3\} is equivalent to \{3,4\} and \{3,5\}, but \{4,2\} is worse than \{2,4\}.

\section{Discussion}
\label{sec:discussion}
While our results are admittedly preliminary, and would require more extensive experimenting to become truly conclusive, we believe they already provide enough interesting evidence to make a case for QoS-aware RSs. 
Next, we discuss   follow-up issues related to  QoS-aware RSs and our experimental results.


\myitem{QoS as ``context''.} The QoS could be considered as an extra dimension in the ``context'' of a user session (e.g., a user watching a YouTube video). For instance, if the user is mobile or in areas with low quality connectivity, QoS awareness could be triggered in the RS. More broadly, 
context-aware RSs have attracted significant attention in the past few years~\cite{villegas2018characterizing,agagu2018context, pagano2016contextual, almutairi2017context, wang2017context, christakopoulou2017recommendation, 
park2019group}. 
The main algorithmic approaches for incorporating contextual information into rating-based RSs are pre-filtering, post-filtering, and modeling~\cite{villegas2018characterizing}. Similar approaches could be considered for the design of QoS-aware RSs. Our work provides initial insights, which can be helpful in the tuning of the pre/post-filtering algorithms, or the development of models amenable to multi-criteria optimization~\cite{musto2017multi, 
hassan2019evaluating
}.


\myitem{Experiments in the wild.} To better understand how the \textit{interest}, \textit{QoR} and \textit{QoS} jointly affect the experience and/or engagement of a user, more experimental work is needed (e.g., through our testbed 
or real-world A/B tests). \myblue{Future experiments should include different types of QoS impairments as well (video quality, latency, bit-rate changes, start-up delay, etc.), whose interplay with QoR and impact on QoE (e.g.,~\cite{plakia2019should}) may differ from the rebuffering impairments that were considered in our initial experiments. \bb{Experiments or measurements over a real streaming service (``in the wild''), e.g., in collaboration with a content-provider, could allow to quantify the QoE with more metrics (e.g., time spent in service, fraction of video watched) that are less subjective than explicit user ratings.}
}

\myitem{Literature on QoS and RSs} There exist works that considered recommendation of web services~\cite{zheng2011qos,chen2013personalized,tang2012location,silic2015prediction,zhang2014temporal}, cloud services~\cite{ding2014combining}, or services at the mobile edge~\cite{wang2017qos}, by taking into account the QoS. Their goal is to predict the QoS, e.g., by using collaborative filtering techniques and taking into account user location~\cite{chen2013personalized,tang2012location}, user clusters~\cite{silic2015prediction}, or time models~\cite{zhang2014temporal}. Then, they inform the user about the available services and the expected QoS. However, this is different from what we propose in this work, i.e., to incorporate QoS-awareness in the RSs themselves. Moreover, we focus on video services, since they are of more interest today, as they dominate traffic~\cite{cisco2018, ericsson2018}, have stringent quality requirements~\cite{conviva2015,ericsson2018}, and make mobile networks struggle in the face of congestion~\cite{cisco2018}.

\section{Conclusion}\label{sec:conclusion}
As first class citizens in the Internet ecosystem, content RSs should increasingly account for aspects intrinsic to the way the data is delivered, including QoS impairments that can be caused by the network state. In this paper, we reported results on how this very foundational aspect of content delivery can impact RSs. We envision that the experimental results presented in this paper are a first step towards embracing QoS into recommendations, to jointly improve the level of satisfaction of  users, content providers and networks.


%

%
\bibliographystyle{ACM-Reference-Format}
\bibliography{references}

%



\end{document}